# EXCEDE Technology Development III: First Vacuum Tests.

Ruslan Belikov[*a], Julien Lozi[b], Eugene Pluzhnik[a], Troy T. Hix[c], Eduardo Bendek[a], Sandrine J. Thomas[a], Dana H. Lynch[a], Roger Mihara[c], J. Wes Irwin[c], Alan L. Duncan[b], Thomas P. Greene[a], Olivier Guyon[b], Richard L. Kendrick[c], Eric H. Smith[c], Fred C. Witteborn[a], Glenn Schneider[b]
[a]NASA Ames Research Center, Moffett Field, CA;
[b]University of Arizona, Tucson, AZ;
[c]Lockheed Martin Space Systems Company, Palo Alto, CA.

## ABSTRACT

This paper is the third in the series on the technology development for the EXCEDE (EXoplanetary Circumstellar Environments and Disk Explorer) mission concept, which in 2011 was selected by NASA's Explorer program for technology development (Category III). EXCEDE is a 0.7m space telescope concept designed to achieve raw contrasts of 1e6 at an inner working angle of 1.2 l/D and 1e7 at 2 l/D and beyond. This will allow it to directly detect and spatially resolve low surface brightness circumstellar debris disks as well as image giant planets as close as in the habitable zones of their host stars. In addition to doing fundamental science on debris disks, EXCEDE will also serve as a technological and scientific precursor for any future exo-Earth imaging mission. EXCEDE uses a Starlight Suppression System (SSS) based on the PIAA coronagraph, enabling aggressive performance.

Previously, we reported on the achievement of our first milestone (demonstration of EXCEDE IWA and contrast in monochromatic light) in air. In this paper, we report on our continuing progress of developing the SSS for EXCEDE, and in particular (a) the reconfiguration of our system into a more flight-like layout, with an upstream deformable mirror and an inverse PIAA system, as well as a LOWFS, and (b) testing this system in a vacuum chamber, including IWA, contrast, and stability performance. The results achieved so far are 2.9e-7 contrast between 1.2-2.0 l/D and 9.7e-8 contrast between 2.0-6.0 l/D in monochromatic light; as well as 1.4e-6 between 2.0-6.0 l/D in a 10% band, all with a PIAA coronagraph operating at an inner working angle of 1.2 l/D. This constitutes better contrast than EXCEDE requirements (in those regions) in monochromatic light, and progress towards requirements in broadband light. Even though this technology development is primarily targeted towards EXCEDE, it is also germane to any exoplanet direct imaging space-based telescopes because of the many challenges common to different coronagraph architectures and mission requirements.

This work was supported in part by the NASA Explorer program and Ames Research Center, University of Arizona, and Lockheed Martin SSC.
**Keywords:** exoplanets, coronagraph, debris disk, EXCEDE, high contrast, IWA, explorer, direct imaging

## 1. INTRODUCTION

In this paper, we report on the continued progress of our technology development effort to mature the Starlight Suppression System (SSS) for the EXoplanetary Circumstellar Environments and Disk Explorer (EXCEDE) mission concept. It is a direct continuation of our first two papers on this effort, published at the previous two years' SPIE conferences [1,2], which contain all the necessary background on our effort, including an overview of EXCEDE, its SSS, and previous results at an the in-air Ames Coronagraph Experiment (ACE) testbed. We repeat a brief overview of these topics in this section and then focus most of the paper on new developments over the past year.

**1.1 Review of the EXCEDE concept.**

EXCEDE is a science-driven technology-pathfinder EX-class Explorer mission concept [3]. It consists of a 0.7m coronagraphic space telescope (see Figure 1) with an unobscured pupil that images in two 20%-wide bands at 0.4 and 0.8 µm. It uses a SSS to suppress starlight to $10^{-6}$ raw contrast between 1.2 and 2 λ/D and $10^{-7}$ raw contrast between 2 and 22 λ/D. With speckle subtraction in post-processing and polarimetric imaging, it is capable of seeing targets down to augmented contrast of $10^{-9}$ (see [3] and Figure 2). The main EXCEDE mission goals are:
1. To characterize the circumstellar environments in habitable zones and assess the potential for habitable planets.
2. To understand the formation, evolution, architectures, and diversity of planetary systems.
3. To develop and demonstrate advanced coronagraphy in space, enabling future exoplanet imaging missions.

---

[*] Ruslan.belikov@nasa.gov, 650-604-0833

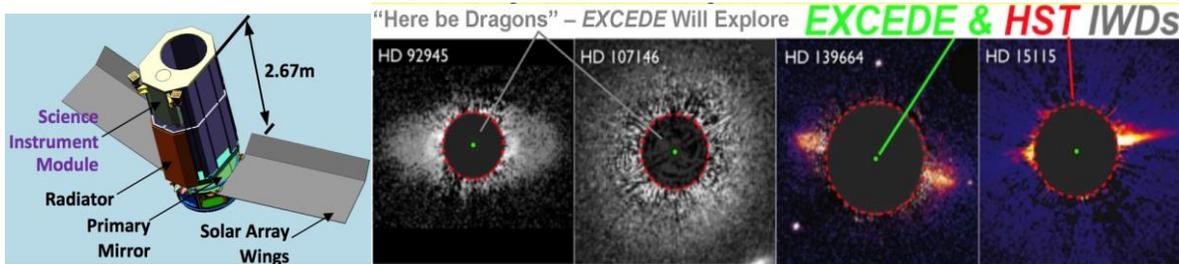

**Figure 1. Left: EXCEDE telescope concept. Right: HST optical images of Circumstellar (CS) disks, with HST/ACS and EXCEDE inner working regions compared. EXCEDE will image ~1000x fainter in contrast and at least 3x closer to their stars and at spatial resolutions comparable to the best JWST will deliver.**

The EXCEDE SSS and its expected performance is shown schematically in Figure 2 and is described more fully in [1,2,3]. It contains three critical components that suppress stellar light arising from 3 different but equally important sources:
1. Fundamental physics, specifically diffraction, which is removed by the coronagraph (blue). EXCEDE SSS is based on the highly efficient Phase Induced Amplitude Apodization coronagraph (see e.g. [4]) which allows it to take full advantage of the smaller aperture, enabling performance of a much larger telescope.
2. Manufacturing limitations such as static and quasi-static wavefront errors, as well as alignment errors, which are removed by the wavefront (WF) control system (green and orange). EXCEDE relies on a WF control system based on focal-plane-based sensing such as the Electric Field Conjugation [5].
3. Environmental disturbances and instabilities such as fast low order errors (e.g. tip/tilt), which are removed by the Low-Order Wavefront Sensor (LOWFS, purple, orange, and yellow) [6].

A fourth critical component is the science camera (which is a two-band Nyquist sampled imaging polarimeter, green in Figure 2).

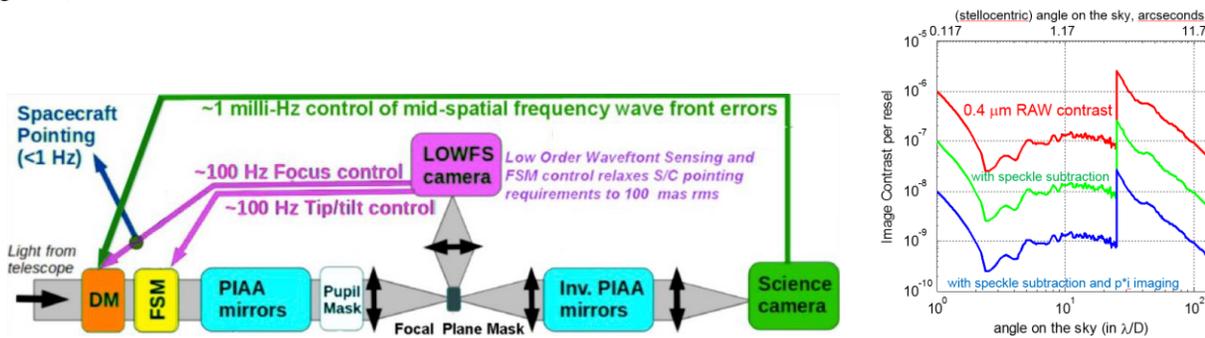

**Figure 2. Left: Schematic of EXCEDE SSS. DM: Deformable mirror; FSM: Fine Steering Mirror; LOWFS: Low Order Wavefront Sensor; PIAA: Phase Induced Amplitude Apodization. Right: Simulated raw contrast after WF control, taking into account expected chromatic aberrations, low order instabilities, known static errors (red curve); contrast after post-processing (green and blue).**

EXCEDE was proposed in response to the Explorer AO in February 2011 by a team led by Dr. Glenn Schneider of University of Arizona (UofA) and is a partnership between UofA, Lockheed-Martin (LM) Corporation, and the NASA Ames Research Center (ARC). EXCEDE was selected as an Explorer Category III investigation ("Scientifically sound investigation requiring further technical development"), receiving technology development funding to mature its proposed SSS to early TRL5.

## 1.2 Summary of progress over the past year

At the past year's SPIE conference, we reported on the following results [2]: (a) 1.8e-7 median contrast between 1.2 and 2.0 $\lambda/D$ and 6.5e-8 between 2.0 and 4.0 $\lambda/D$, in monochromatic light (b) preliminary broadband results, and (c) a successful integration and demonstration of the Low Order Wavefront Sensor [7]. However, these results were obtained with a simplified layout that did not contain the inverse PIAA mirrors and where the DM was downstream instead of upstream of the PIAA mirrors. Also, the results were obtained in an in-air testbed.

This year, there are three key significant developments: (a) we conducted our first vacuum tests; (b) we reconfigured our optical layout to more closely match the EXCEDE one in Figure 2; and (c) we made significant progress in studying broadband performance of our system. We describe these results in what follows. Specifically, section 2 describes our new vacuum testbed; section 3 describes our new layout; 4 describes our results and limiting factors; and finally in section 5 we conclude with a discussion of the significance of the results in the context of other missions, demonstrations, and theoretical limits. In addition, a companion paper [8] contains detailed coverage of our Low Order Wavefront Sensor system (LOWFS) as well as an analysis of our stability and drifts, which were critical to the achievement of our results and critical to the actual mission.

## 2. VACUUM TESTBED

A major new step in technology development taken this year is a vacuum test of an EXCEDE-like system. This involved preparing a vacuum chamber, and making sure that all parts of our system are vacuum-ready. This in particular involved making our science and LOWFS cameras vacuum compatible, procuring vacuum-compatible stages and motors, and installing vacuum-compatible feedthroughs for fibers, cameras, motors, and the DM. Significantly, our tests represent a successful vacuum high contrast demonstration with a Boston Micromachines 1K DM.

### 2.1 Vacuum Chamber

The test chamber used for the results in this paper is located at The Lockheed Martin (LM) Palo Alto, CA facility. This thermal vacuum 'Metrology' (MET) chamber is unique for its shape and size, operating environment, and the facility in which it is embedded. The 8 ft (2.4 m) wide chamber is tall enough to allow relatively comfortable access to the entire length of the 4 x 20 x 2 ft (1.2 x 6.1 x 0.6 m) optical table that is integrated to the chamber through vibration isolators located external to the chamber. The payload entrance to the chamber is embedded in a 20 ft (6.1 m) wide 40 ft (12.2 m) long class 1000 cleanroom with a flow bench/ESD station, 10 ft (3.0 m) optical bench, pneumatic movable scissor jack for breadboard insertion, and a 2 ton (1815 kg) gantry. The vacuum environment within the chamber is produced by an Edwards dry roughing pump, a cryogenic pump, a turbo molecular pump, and a liquid nitrogen getter plate. The chamber has eight, fourteen-inch ports located on each side of the chamber for data I/O, or optical access as required. In addition to the chamber, LM has made available a vacuum compatible baseplate for the development and test of a vacuum-compatible SSS (for this test, however, the apparatus was assembled directly to the optical bench) and has installed feed-throughs for all system cables (including those associated with the deformable mirror) and has activated/installed active cooling systems for the optical bench and cameras. Data logging systems are in place such that environmental data (pressure, temperature, acoustic vibration) can be recorded and transferred.

We conducted three vacuum tests so far and in this paper will focus on the final test. For this test, the chamber door was closed and we began pumping on 2/24/14. The optical bench was passively temperature controlled (the control loop was open) and the getter plate was not used – in this way the test chamber could be operated without 24/7 staffing. For the first half of the 66 day vacuum test, the cryo pump and turbo molecular pump (TMP) were both used. There were a few periodic blips in pressure as trapped gasses in the system found their way out (day 7 and day 18 were notable). By day 32 the system was settling at the single digit uTorr level (although our experiments only required ~100µTorr which was achieved in less than 1 day.) At this point the cryo pump was turned off and the chamber was run on the TMP alone. In this way we minimized mechanical vibrations coupling into the system since the cryo pump contributed a majority of the noise in the pumping system. Immediately after the switch the vacuum retreated to the 10 uTorr level and slowly fell below 10 uTorr over the remainder of the test. We took the chamber up to air and opened the door on 5/1/14. See Figure 3 for a plot of the recorded chamber pressure during the test.

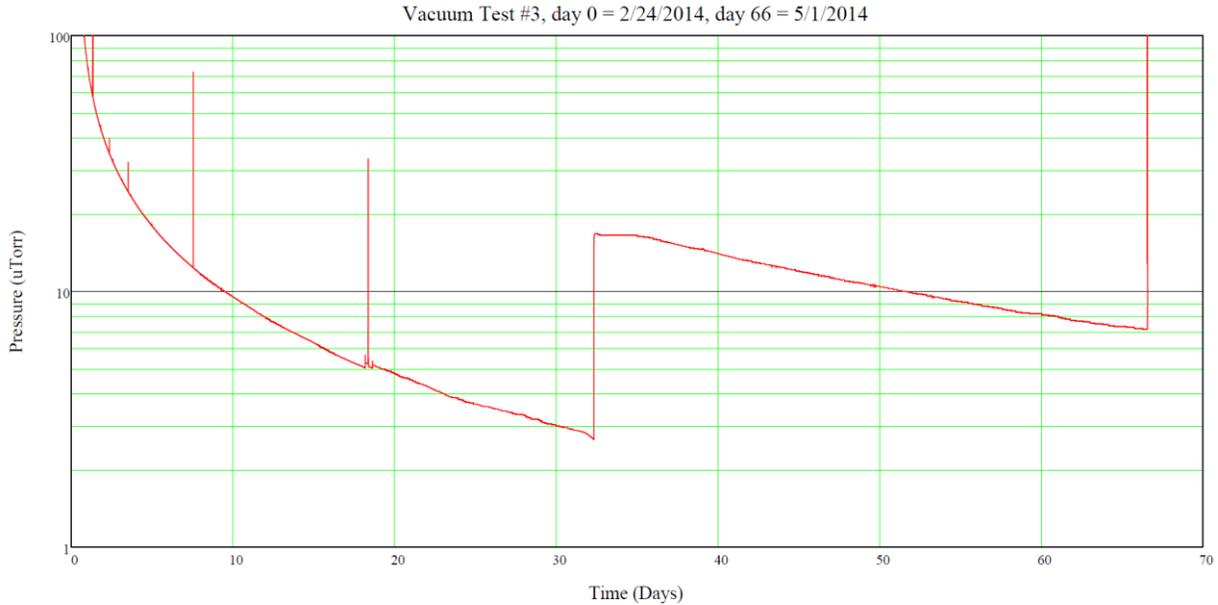

**Figure 3. Chamber Pressure**

.

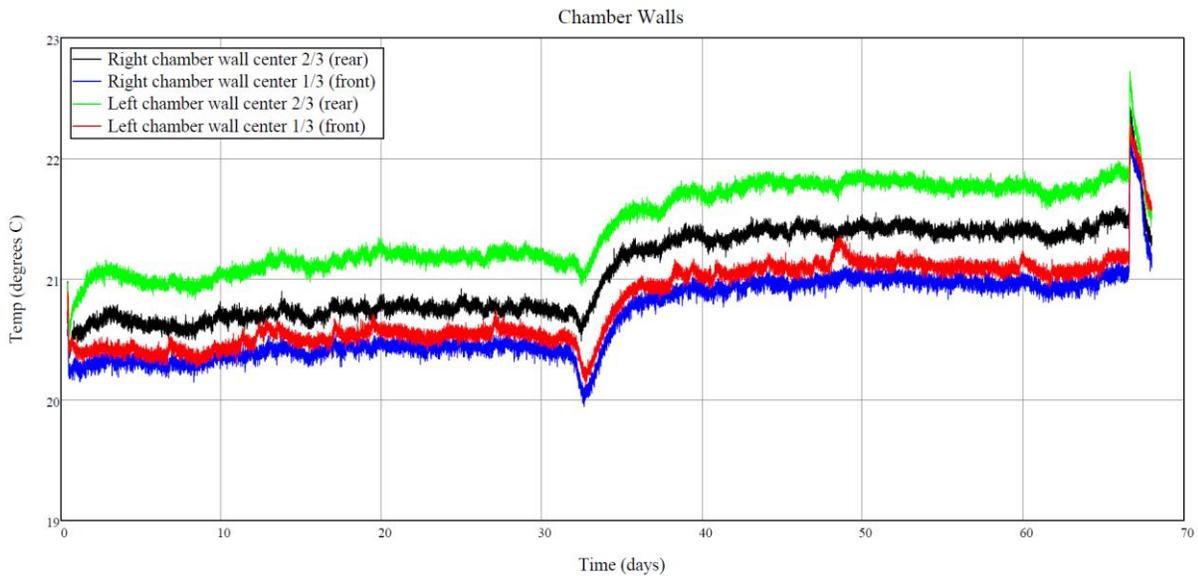

**Figure 4. Chamber Wall Temperature**

Temperature sensors (un-calibrated) were placed at various locations to monitor conditions on the test system and the chamber. Four sensors were placed on the walls of the chamber at about chest height – two equally spaced along the right chamber wall and two equally spaced along the left chamber wall. These sensors show that the short term (1 day) stability of the wall temperature was less than 0.1 degree C. There was a gradient of about 1 degree C from wall to wall; however some of this may be due to calibration differences which were unaccounted for. The long term temperature trace shows a jump at day 32 when the vacuum pump was switch, as would be expected. Sensors on the floor and table show similar behavior. See Figure 4 for a plot of the temperature traces from the wall sensors.

Passive (open loop) cooling systems were installed for the two cameras in the optical system. A recirculating cooling system using Novec fluid was plumed into the chamber and run through cooling plates attached to the cameras. The system needed to accommodate moving cameras while minimizing vibrational coupling between the fluid flow and the optical system. For the science camera (made by QSI) the vibration was not a significant issue and the cooling plate was attached directly to the camera body. For the LOWFS camera (made by Imperx), the vibration issues are much more significant. Thus, in the LOWFS camera case, the cooling plate was mechanically attached using flexible copper rope.

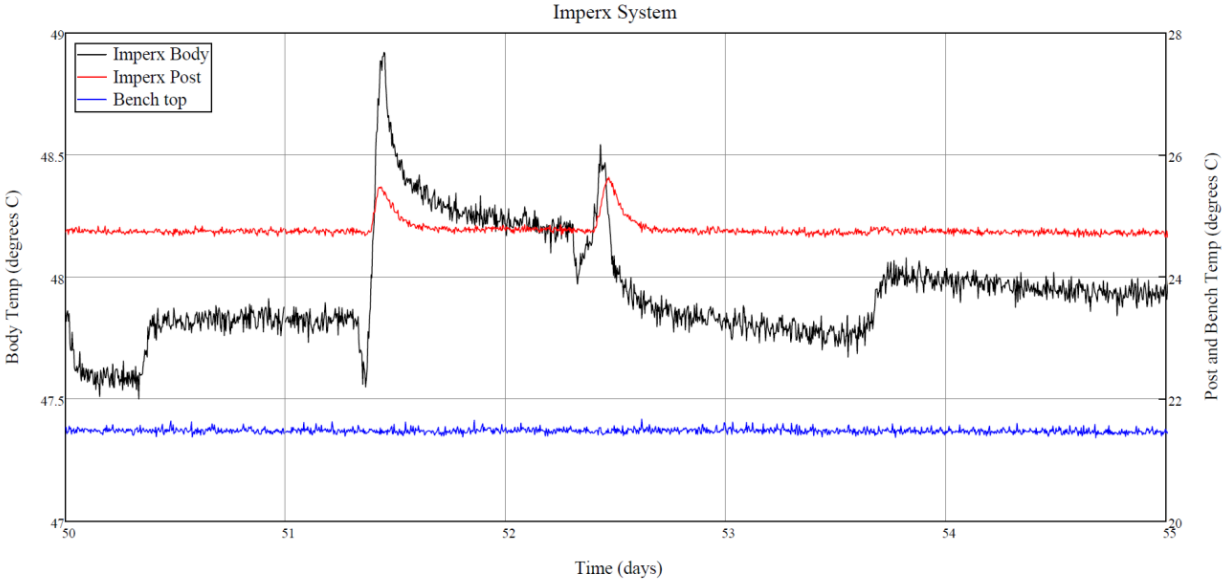

**Figure 5. Imperex Temperature Sensors**

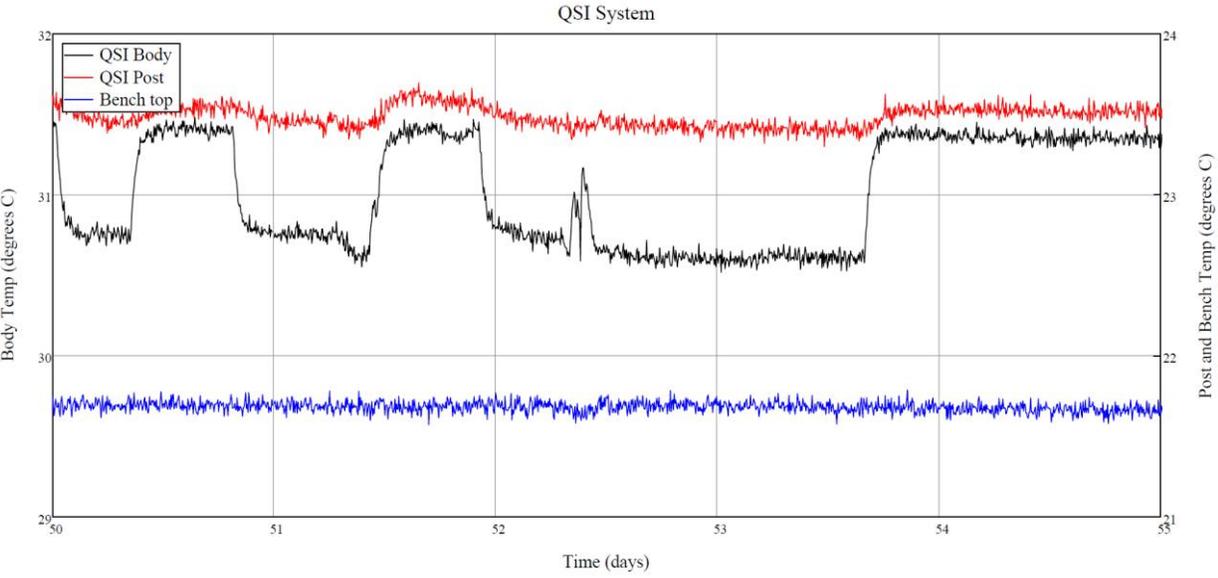

**Figure 6. QSI Temperature Sensors**

To monitor the effectiveness of the camera cooling systems, temperature sensors were attached to the camera body, to the camera post supporting the body, and to the optical bench surface nearby. Thus, in addition to the camera temperature, the heat leakage from the camera to the post and subsequently into the optical bench could be monitored. Figures 5 and 6 show the temperature sensor output for the Imperx system and the QSI system respectively. Note that the temperature of the body is read off the left axis while the temperature of the post and table are read off the right axis. In these figures we see a 23 (9) degrees C delta between the post and body on the Imperx (QSI) camera, while the post/table delta was around 4(2) degrees C. The most important result was the fact that these systems reached equilibrium and also allowed the cameras to work in their specified temperature limits.

Even though the bench is massive and could be used as a heat sink, it was preferable that we minimize the amount of heat dumped into the optical bench. In this way we minimize the possibility of misalignments of nearby optics due to thermal expansion of the table surface. Note that there is no measureable rise in the temperature of the table surface, implying that most of the heat went out through the cooling fluid (at least enough to minimize the rise in temperature of the table surface).

## 3. NEW EXCEDE-LIKE LAYOUT.

For the tests described in this paper, we used a new, updated optical layout that is much closer to the EXCEDE design than previous years' tests. (See [9] for more detail.) The most significant changes were: (a) the deformable mirror is now upstream, rather than downstream of the PIAA system; and (b) an inverse PIAA system is used. This combination allows a much wider field of view and outer working angle of the deformable mirror, but is more challenging to align and perform wavefront control.

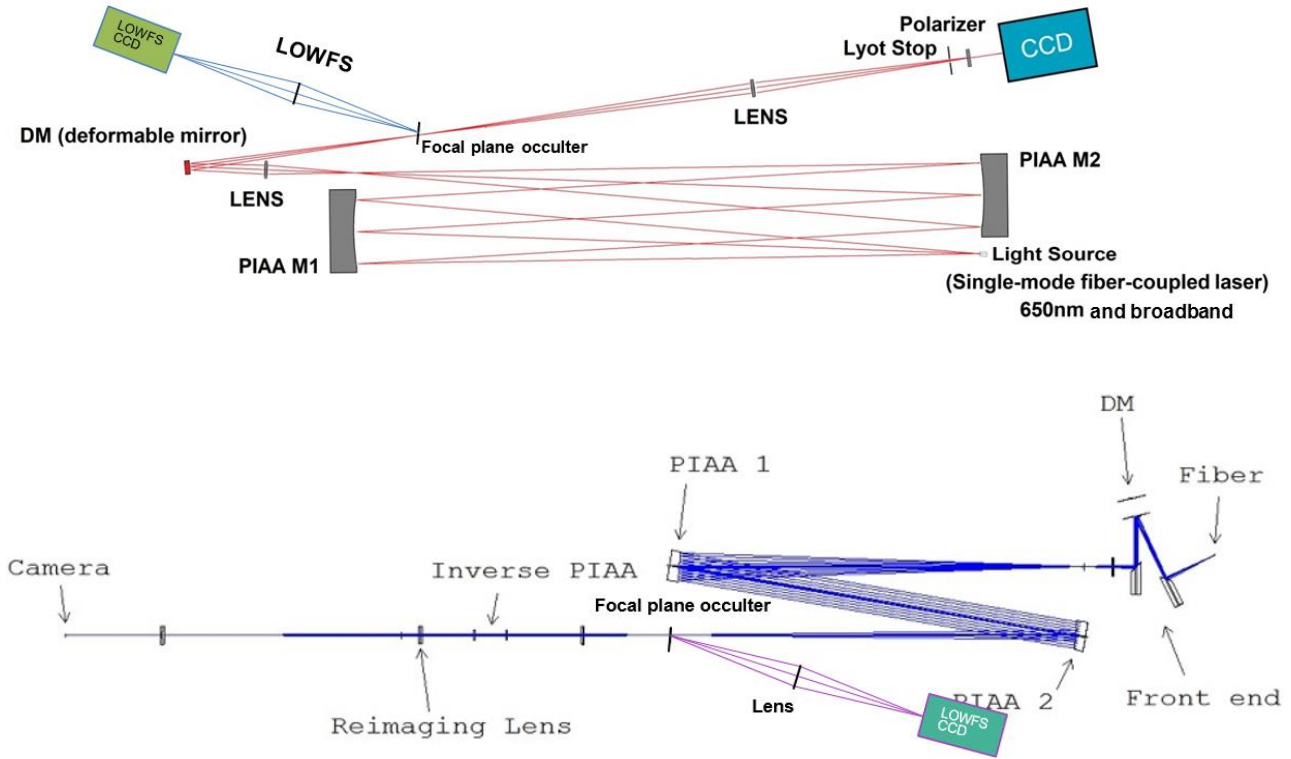

**Figure 7. Top: Old layout used for previous year's tests (in air). Bottom: New Zemax layout used for tests used for the demonstrations in this paper.**

The input to our system is a laser fed by a fiber into the vacuum chamber. We have a number of fibers in a bundle which allows us the flexibility to test different kinds of fibers at the source and match the fiber to the wavelength range used.

As in the real EXCEDE design, the deformable mirror is conjugated to the PIAA system. However, unlike the EXCEDE design, the mirror is a 1K rather than a 2K device. This was not an important difference for our current technology development goals, which are to demonstrate high contrast in broadband light primarily in the vicinity of the inner working angle, which is the most challenging part of the science image plane.

We did not yet make an occulter mask matched to the layout on the bottom of Figure 7 and simply used the same occulter (between the PIAA and inverse PIAA system) as was used in [2] and on the top of Figure 7, i.e. a chrome mask on glass. Because the f-number in the occulter plane is much faster on the bottom of Figure 7 than the top, our occulter was oversized. This did not prevent us from blocking the star or reaching high contrasts, but it did create certain issues described in the next section and contributed to the limiting factors.

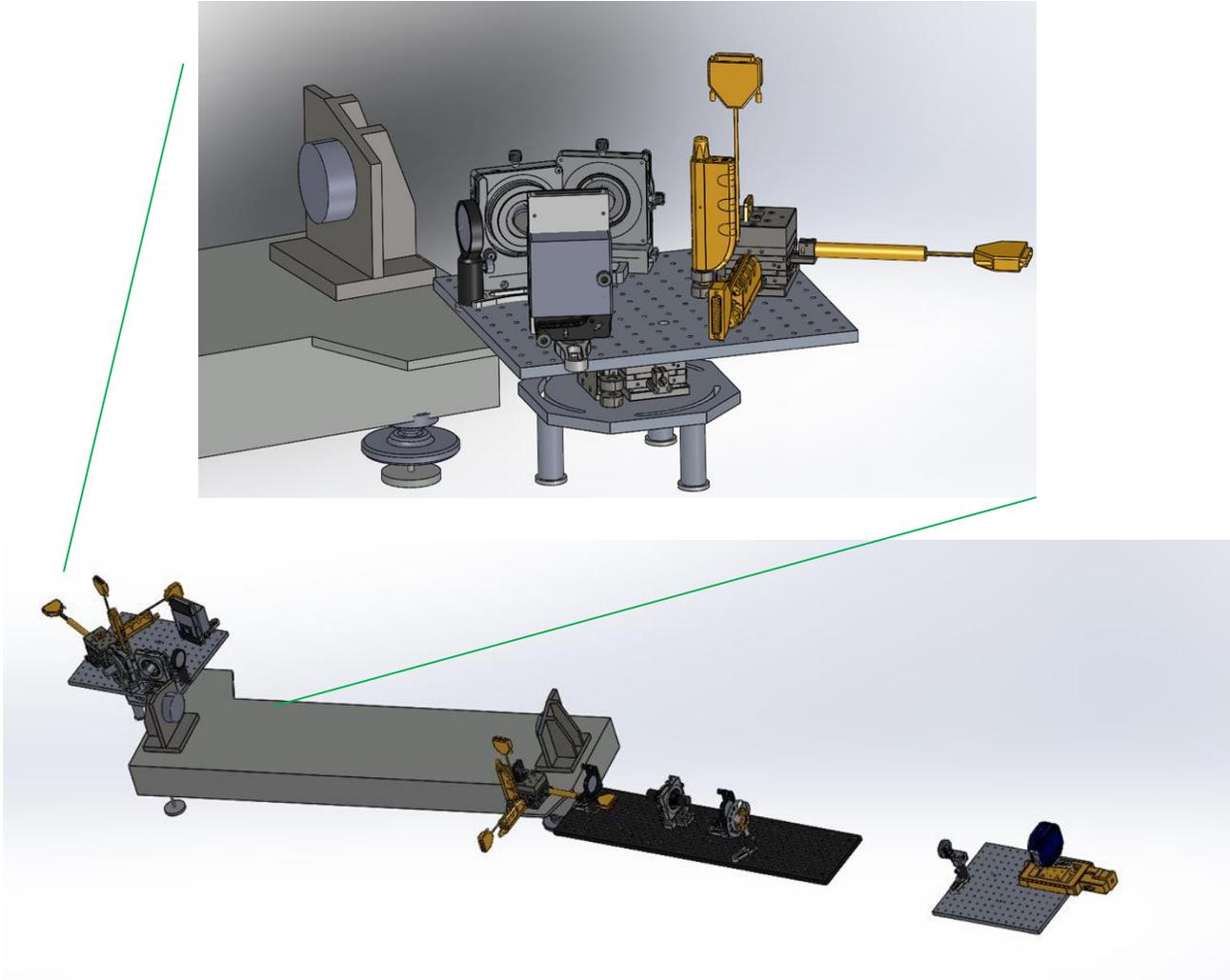

**Figure 8. CAD layout of the vacuum system**

Our forward PIAA system consisted of two 75mm active diameter mirrors made by Axsys, and our reverse system consisted of 16mm active diameter lenses. Both sets of optics convert from a top-hat illumination profile to a prolate spheroidal (or vice versa) in the middle of the beam, but have different profiles at the edges of the beam (because they were designed for different systems). Nevertheless, the degradation in Strehl seems to be negligible as described in the next section. Because very little starlight illuminates the inverse PIAA system (being blocked by the focal planet occulter), high quality mirrors are not required for the inverse PIAA system and we used lenses both for ease of alignment and to save physical space on the bench. We have not yet fully characterized their impact on broadband performance and we may switch to mirrors if they are determined to be a limiting factor. However, they do not seem to be our current limiting factor.

As before, we are using a polarizer in front of our CCD camera, but we have not yet checked whether it is indeed necessary. In the actual EXCEDE layout, the polarizer is replaced by a filter wheel with a Wollaston prism, which we plan to eventually implemented on our testbed.

Figure 8 shows a CAD layout of our vacuum system, implementing the Zemax layout from Figure 7, and demonstrates the actual configuration of our hardware. The top of the figure shows our "front end" which is placed on a separate plate on a stage for ease of alignment and better stability. The PIAA mirrors are mounted on their own separate bench aligned to the front end, followed by the rest of the system.

## 4. RESULTS AND LIMITING FACTORS

We have performed a total of 3 vacuum tests so far, all of which were very successful. The first vacuum test was focused on testing system stability, Strehl, and alignment (without the inverse PIAA system), with results sufficient to meet EXCEDE contrast and IWA requirements (see [2] for more details on stability of our system). The second test focused on testing of the system Strehl and stability with the inverse PIAA system and revealed that the act of pumping misaligns the inverse PIAA too much, which resulted in us implementing a special alignment stage for it. The third test focused on full system wavefront control in monochromatic light as well as performance measurements in both monochromatic and broadband light.

### 4.1 Characterizations and calibrations

We briefly describe our characterizations and calibrations. Most of these are implemented by an automated LabVIEW interface, and our protocol involves checking the IWA and photometric calibrations both before and after any wavefront control run.

#### 4.1.1 Strehl ratio

Although high contrast can in principle be achieved even on a system with a low Strehl ratio, it is always good to maximize it because it makes both the wavefront control and planet detection easier. The Strehl ratio of a PIAA system is strongly dependent on alignment. We computed our Strehl ratio to be 0.9 (on-axis) at the intermediate focal plane between the PIAA optics. At the final focal plane, the Strehl ratio is 0.85 (peak). This indicates acceptable alignment, and is in fact better than during our previous in-air tests [1,2]. However, the Strehl degrades to about 50% when for sources that are ~15 l/D off-axis. This is expected given that we have not yet matched the shape and the position of the inverse PIAA system to the forward system, and may affect wavefront control at large off-axis angles. Nevertheless, this does not (yet) appear to be our limiting factor.

#### 4.1.2 Photometry calibrations

We attempted and cross-validated several photometry calibrations in order to accurately compute the contrast of our results. This included: (a) creating speckles with decreasing brightness with a DM, measuring the contrast of the brightest speckle with the focal plane occulter out, putting the focal plane occulter in and creating a dimmer speckle, then measuring the ratio of the dimmer speckle with respect to the bright speckle. Then the dimmer speckle can be used as a reference for contrast in the dark zone. (b) Using natural speckles far from the optical axis. We found that the latter method works better because the speckles in (a) depend on whether the focal plane occulter is in or out.

#### 4.1.3 Inner Working Angle and plate scale calibrations

We define the IWA as the angle of the source at which the total coronagraphic throughput is 50% of maximum (i.e. a far off-axis source). Plate scale is the number of pixels per l/D on the science camera. Both of these parameters can be computed with models if all the system parameters (distances between components, focal lengths, etc.) are known accurately and precisely. However, this is often not the case. We opt for a direct measurement of the plate scale and inner working angle. Our fiber and focal plane mask are placed on very precise motorized stages allowing displacing them by precise and known amounts. We measure the plate scale by moving the fiber with the occulter out and analyzing the motion of the PSF on the science CCD. (This also allows us to measure Strehl.) We set the IWA very precisely by displacing the fiber to a desired IWA, and then moving the focal plane mask to a position where the total measured throughput at the science camera is 50% of maximum. One also has to be careful about the fact that wavefront control can change the photocenter of the on-axis PSF and therefore affect the IWA. This can be compensated either by constraining the wavefront control algorithm to keep the same IWA, or simply to compensate for this by setting a more aggressive IWA than desired before starting wavefront control, and then verifying the IWA after wavefront control. We opted for the second option. In all the results below, the IWA was approximately 1.2 l/D.

### 4.2 Monochromatic results

After performing the calibrations and characterizations described above, we ran our wavefront control algorithm to create a dark zone in a C-shaped region of interest as shown in Figure 9. We used the same "Speckle Nulling" wavefront control algorithm as described in [1,2]. Although this algorithm is slow, it does not require a well-calibrated model of the system and is a good way of checking initial performance. We expect to apply faster algorithms such as EFC once limiting factors are eliminated and once we calibrate our system model

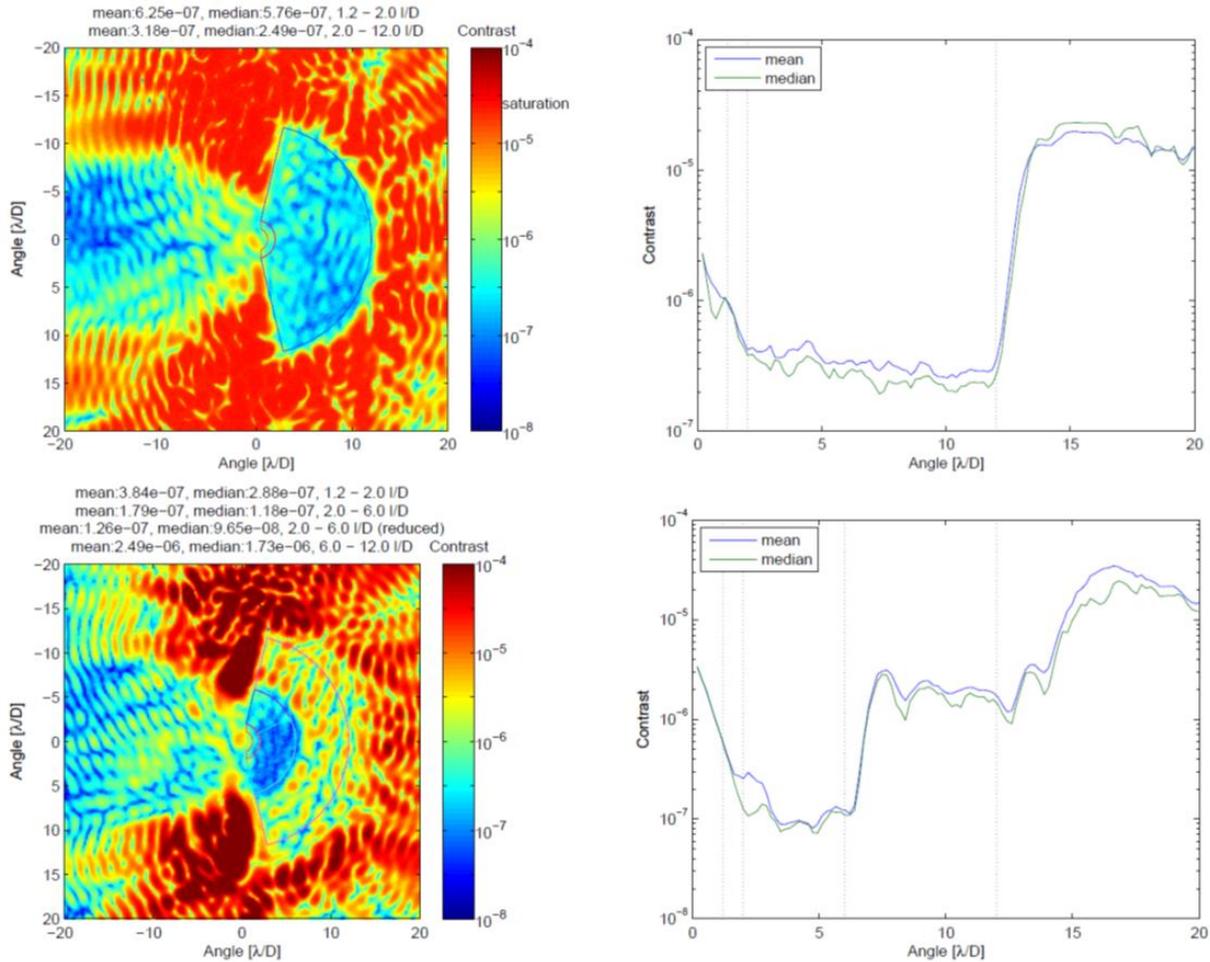

**Figure 9. Top: Large OWA case. Median contrasts are: 5.8e-7 between 1.2 - 2.0 l/D and 2.5e-7 between 2.0 - 12 l/D. Bottom: narrow OWA case. Median contrasts are: 2.9e-7 between 1.2-2.0 l/D and 9.7e-8 between 2.0-6.0 l/D.**

We first created a dark zone between 1.2 and 12 l/D, close to the Nyquist limit of the deformable mirror. Median contrasts were 5.8e-7 between 1.2-2.0 l/D and 2.5e-7 between 2.0-12 l/D. We then reduced the size of the region of interest by lowering the outer working angle from 12 to 6 (bottom of Figure 9), which improved the contrast to 2.9 between 1.2-2.0 l/D and 1.2e-7 between 2.0-6.0 l/D. Therefore, the limiting factor for the larger zone may be the algorithm or insufficient number of iterations. (Once EFC is used, this may improve.) However, even the reduced dark zone is already larger than the one demonstrated with our previous layout [2], demonstrating the power of the new layout.

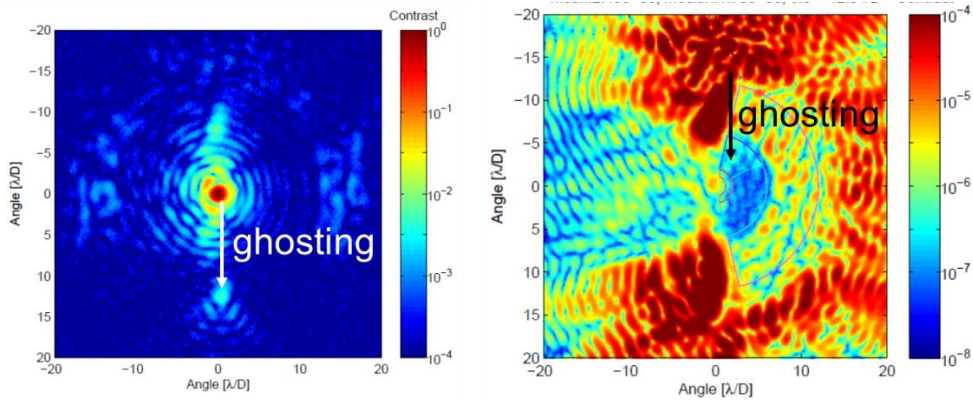

**Figure 10. Left: very short exposure image with the focal plane mask out, showing a ghost. Right: the same ghosting affects the dark zone and is believed to be our limiting factor. This limiting factor is being addressed by an anti-reflective coating on the focal plane mask**

We traced the limiting factor to be a ghost created by a reflection from the surfaces of our focal plane mask. This ghost can be seen as speckles in the top third of the dark zone region. If we exclude that top third, the contrast in the remainder of the region is better than 1e-7. An anti-reflective coating on the focal plane mask is expected to eliminate this limiting factor in the next vacuum test.

### 4.3 Broadband results

Our initial goals focused on demonstrating a system closer to the EXCEDE system for the simpler case of monochromatic light. Therefore, we did not yet optimize our layout for broadband light (for example, we used lenses in several places rather than mirrors to speed up alignment and testing). Nevertheless, we tested the performance of our system in broadband light and the results are below. For these tests, we first created a dark zone with our monochromatic laser and then switched to a whilte light (supercontinuum) laser. Our white light laser contained a filter allowing us to select a band with an arbitrary width and center wavelength. We set the center wavelength to match the monochromatic wavelength (655nm) and varied the bandwidth from 2nm to 128nm by powers of 2. The results are shown in Figure 11.

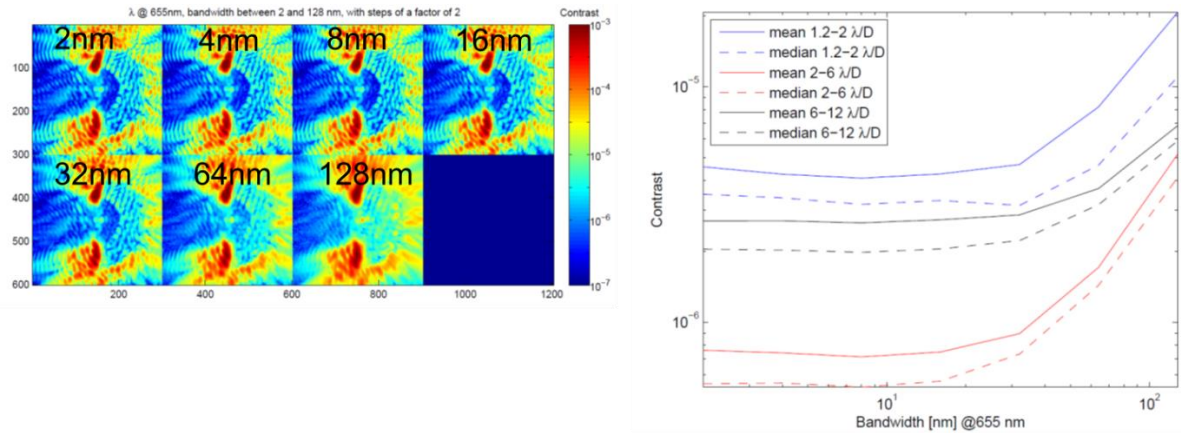

**Figure 11. Preliminary broadband results. As measured between 2-6 l/D, contrast is 7.3e-7, 1.4e-6, and 4.1e-6 for a ~5, 10, and 20% bands, respectively. These are likely limited by the fact that our current system uses chromatic lenses for simplicity to make faster progress in monochromatic light as a first step. We will switch to achromatic components once monochromatic limiting factors are eliminated.**

Despite having chromatic components on our system, our results already demonstrate reasonable achromaticity across ~30nm, or 5%. However, note that even at 2nm bandwidth, which is almost monochromatic, the contrast is not as good as with a laser. This was determined to be not due to the 2nm bandwidth, but due to the fact that the laser had a long coherence length and thus could better correct ghosts. (Because short coherence length lasers were found to perform similarly to the white light laser set at a 2nm bandwidth.) We suspect that the focal plane mask ghost is the dominant issue and thus broadband contrast is expected to improve by using a focal plane mask with an anti-reflective coating in the next vacuum test.

## 5. SIGNIFICANCE AND CONTEXT OF RESULTS

The significance of the technology development for EXCEDE and the results presented goes well beyond EXCEDE because we are probing an area of performance space at very aggressive inner working angles, something that is not usually considered for larger missions such as WFIRST / AFTA [10] or Exo-C [11]. This is illustrated on Figure 12, which shows a design space for coronagraphs where each point represents a coronagraph with a particular IWA and raw contrast performance. The blue circle labeled as "ACE/LM (2014)" represents the performance (in monochromatic light) described here. The point labeled as "HCIT2 (2014)" represents performance achieved with a coronagraph at NASA JPL High Contrast Imaging Testbed 2 (HCIT2) with a PIAA coronagraph [12] (a collaboration we are part of outside the scope of EXCEDE). These two efforts push performance in two complementary directions as shown by black arrows: improving inner working angle is more significant for small telescopes which may not benefit from very deep contrasts because they would require very long integration times, and where aggressive inner working angle help mitigate their small size. On the other hand, large telescopes can benefit from deeper contrasts, but require greater stability to achieve aggressive inner working angles. Although the two directions are orthogonal, they both push the same boundary as indicated by the dotted line. The dotted blue line represents the approximate limit of our designs in the presence of 3e-3 l/D tip/tilt errors, which is not far from the actual tip/tilt errors on the two testbeds. Note that in the stability-limited regime, ~1e-7 contrast with 1.2 l/D IWA is "equivalent" to 1e-9 contrast with 1.9 l/D IWA, in the sense that both lie on the same trade surface.

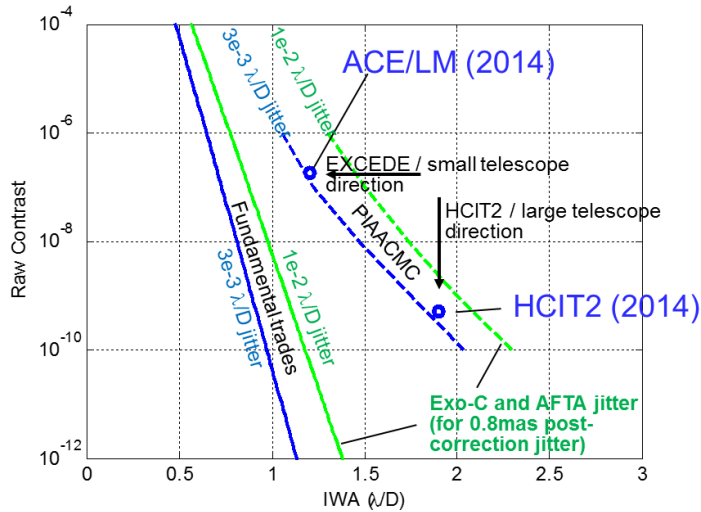

**Figure 12. Contrast - IWA trade space, showing our results (ACE/LM 2014) in the context of other missions and demonstrations.**

The solid lines show fundamental limits for any coronagraph [13] and represent potential performance that can be achieved in the presence of shown telescope jitter with better coronagraph designs. Note that the green lines represent the trade surface corresponding to the level of jitter specified for the WFIRST/AFTA and Exo-C mission concepts. Pushing coronagraph designs and performance past the dotted lines closer to the solid lines for EXCEDE and the resulting lessons learned can help these other missions as well.

## 6. CONCLUSIONS

Technology development is advancing for the EXCEDE mission, an Explorer mission concept designed to directly image debris disks and capable of seeing down to 10 zodis in the habitable zones of many nearby stars. The main results reported here are a start of vacuum testing of an EXCEDE-like optical configuration. The results achieved so far are 2.9e-7 contrast between 1.2-2.0 l/D and 9.7e-8 contrast between 2.0-6.0 l/D in monochromatic light; as well as 1.4e-6 between 2.0-6.0 l/D in a 10% band, all with a PIAA coronagraph operating at an inner working angle of 1.2 l/D. This constitutes better contrast than EXCEDE requirements (in those regions) in monochromatic light, and progress towards requirements in broadband light. In addition, our Low Order Wavefront Sensor (LOWFS) was successfully demonstrated in vacuum, which is a critical component of the EXCEDE Starlight Suppression System and enables the stability required for aggressive inner working angles, and the Boston Micromachines 1K deformable mirror has been shown to successfully operate in vacuum and achieve high contrasts. Finally, this work is also relevant to other missions and other coronagraphs because it explores fundamental physics limits that apply to all coronagraphs and raises the TRL of technologies necessary for operation at aggressive IWAs.

## ACKNOWLEDGEMENTS

This work was supported in part by the National Aeronautics and Space Administration's Ames Research Center, as well as the NASA Explorer program and the Technology Development for Exoplanet Missions (TDEM) program through solicitation NNH09ZDA001N-TDEM at NASA's Science Mission Directorate. It was carried out at the NASA Ames Research Center and Lockheed Martin ATC. Any opinions, findings, and conclusions or recommendations expressed in this article are those of the authors and do not necessarily reflect the views of the National Aeronautics and Space Administration or LM.

## REFERENCES

1. Belikov, R., Pluzhnik, E., Witteborn, F.C., Greene, T.P., Lynch, D.H., Zell, P.T., Schneider, G., Guyon, O., Tenerelli, D., "EXCEDE Technology Development I: First demonstrations of high contrast at 1.2 λ/D for an Explorer space telescope mission," *Proc SPIE 8442-6* (2012).
2. Belikov, R., Bendek, E., Greene, T.P., Guyon, O., Lozi, J., Lynch, D. H., Newman, K. E., Pluzhnik, E., Schneider, G., Tenerelli, D., Thomas, S.J., Witteborn, F. C., "EXCEDE Technology Development II: Demonstration of High Contrast at 1.2 l/D and preliminary broadband results," *Proc SPIE 8864-31* (2013).
3. Guyon, O., Schneider, G. H., Belikov, R., Tenerelli, D.J., "The Exoplanetary Circumstellar Environments and Disk Explorer (EXCEDE)," *Proc. SPIE 8442* (2012).
4. Guyon, O., Martinache, F., Belikov, R., Soummer, R., "High Performance PIAA Coronagraphy with Complex Amplitude Focal Plane Masks," *ApJS* Vol. 190, issue 2, pp. 220-232, 2010.
5. Give'on A., Belikov R., Shaklan S., and Kasdin J., "Closed loop, DM diversity-based, wavefront correction algorithm for high contrast imaging systems," *Optics Express*, Vol. 15, Iss. 19, pp. 12338-12343, 09/2007.
6. Guyon, O., Matsuo, T., Angel, R., "Coronagraphic Low-Order WaveFront Sensor: Principle and Application to a Phase-Induced Amplitude Coronagraph," *ApJ*, Vol 693, Issue 1., pp. 75-84 (2009).
7. Lozi, J., Belikov, R., Schneider, G., Guyon, O., Pluzhnik, E., Thomas, S. J., Martinache F., "Experimental Study of a Low-Order Wavefront Sensor for the High-Contrast Coronagraphic Imager EXCEDE," *Proc SPIE 8864-23* (2013).
8. Lozi, J., Belikov, R., Thomas, S. J., Pluzhnik, E., Bendek, E. A., Guyon, O., Schneider, G. H., "Experimental study of a low-order wavefront sensor for high-contrast coronagraphic imagers: results in air and in vacuum," *Proc SPIE 9143-66* (2014).
9. Bendek, E., Lozi, J., Schneider, G., Thomas, S., Pluhznik, E., Newman, K., Lynch, D., Belikov, R., "Opto-mechanical design of the vacuum compatible EXCEDE's mission testbed," Proc. SPIE 9143-204 (2014).
10. http://wfirst.gsfc.nasa.gov/
11. http://exep.jpl.nasa.gov/stdt/exoc/
12. Lawson, P. R., Belikov, R., Cash, W., Clampin, M., Glassman, T., Guyon, O., Kasdin, N. J., Kern, B. D., Lyon, R., Mawet, D., Moody, D., Samuele, R., Serabyn, E., Sirbu, D., Trauger, J., "Survey of Experimental Results in High-Contrast Imaging for Future Exoplanet Missions," *Proc SPIE 8864-50* (2013).
13. http://exep.jpl.nasa.gov/exopag/exopag10/presentations/Belikov_ExoPAG10_coronagraph_fundamental_limits_v2.pdf.